\documentclass[superscriptaddress, aps, prb, twocolumn, english, nobalancelastpage]{revtex4-2}
\usepackage{amssymb}
\usepackage{enumerate}
\usepackage{graphicx}
\usepackage[dvipsnames]{xcolor}
\usepackage{dcolumn}
\usepackage{bm}
\usepackage{oplotsymbl}
\usepackage{hyperref}
\hypersetup{pdfencoding=auto,colorlinks=true,linkcolor=blue,citecolor=blue}
\usepackage{amsmath, dsfont, complexity}
\usepackage{braket}
\usepackage{multirow}
\usepackage[english]{babel}
\usepackage[T1]{fontenc}

\graphicspath{{./}{./Figures/}}

\newcommand{\eightpointedstar}[1]{%
  \makebox[1.5ex][l]{
    \tikz[baseline=-0.6ex, scale=0.12] \fill[#1]
      (0:1) -- (22.5:0.4) -- (45:1) -- (67.5:0.4)
      -- (90:1) -- (112.5:0.4) -- (135:1) -- (157.5:0.4)
      -- (180:1) -- (202.5:0.4) -- (225:1) -- (247.5:0.4)
      -- (270:1) -- (292.5:0.4) -- (315:1) -- (337.5:0.4) -- cycle;
  }%
}

\newcommand{\Figref}[1]{Fig.~\ref{#1}}
\newcommand{\Eqref}[1]{Eq.~(\ref{#1})}
\newcommand{\Secref}[1]{Sec.~\ref{#1}}
\newcommand{\Appref}[1]{App.~\ref{#1}}

\newcommand{\dipc}{Donostia International Physics Center (DIPC), Manuel Lardizabal Pasealekua 4, E-20018 Donostia, Basque Country}
\newcommand{\ikerbasque}{IKERBASQUE, Basque Foundation for Science, Euskadi Plaza 5, E-48009 Bilbao, Basque Country}

\begin{document}

\title{Identification and optimization of accurate spin models for Fermi-Hubbard ladders using matrix product states}

\author{Andoni Agirre}
\email{andoni.agirre@dipc.org}
\affiliation{\dipc}
\affiliation{Department of PMAS: Physics, Chemistry and Technology, University of the Basque Country (UPV/EHU), Manuel Lardizabal Pasealekua 3, E-20018 Donostia, Basque Country}

\author{Thomas Frederiksen}
\affiliation{\dipc}
\affiliation{\ikerbasque}

\author{Géza Giedke}%
\affiliation{\dipc}
\affiliation{\ikerbasque}

\author{Tobias Gra{\ss}}
\affiliation{\dipc}
\affiliation{\ikerbasque}

\date{\today}

\begin{abstract}
Open-shell nanographenes offer a controlled setting to study correlated magnetism emerging from $\pi$-electron systems. Here, we study non-bipartite Fermi-Hubbard ladders describing oligo(indenoindene) molecules. These feature a gapped, weakly dispersing manifold of quasi-zero modes in their single-particle spectra, and we show that their low-energy properties can be effectively mapped onto an interacting set of spin-1/2 degrees of freedom. Using density matrix renormalization group (DMRG) simulations of the full Fermi-Hubbard model, we obtain their excitation spectra, entanglement profiles, and spin–spin correlations. We then construct optimized delocalized fermionic modes that act as emergent spins and demonstrate that their interactions are well described by a frustrated $J_1$--$J_2$ Heisenberg chain. This effective description clarifies how spin degrees of freedom arise and interact in non-bipartite nanographene ladders, providing a compact and accurate representation of their correlated behavior.
\end{abstract}

\maketitle

\section{Introduction}
Spins are the paradigmatic quantum degree of freedom, particularly well suited for studying many-body quantum physics and quantum technologies. Their suitability is strengthened by the existence of highly developed control techniques such as electron spin resonance, nuclear magnetic resonance, and spin-photon interfaces, as well as their potential for strong isolation from uncontrolled degrees of freedom, which results in long coherence times.
Beyond this, spin physics can give rise to a plethora of fascinating effects such as frustration, quantum phase transitions, spin-fractionalization, or topological excitations~\cite{Auerbach1994,McClarty2022}.
Graphene, which is usually not magnetic, may not come to mind at first as a host material for spins; however, a rich variety of spin physics emerges from graphene-based nanostructures~\cite{Yazyev2010,deOteyza2022}. 
Weak spin-orbit and hyperfine interaction, combined with the flexibility in engineering atomically precise nanostructures, including the placement of heteroatoms by on-surface synthesis techniques, make these structures a versatile and promising testbed for exploring spin physics at the nanoscale. 
Recent advances have enabled the bottom-up fabrication and low-temperature scanning probe characterization of a variety of open-shell graphene-based structures \cite{deOteyza2022}, including single-spin localization and spin-spin interactions in graphene nanoribbon junctions \cite{LiSaCo.19.Singlespinlocalization,WaSaCa.22.MagneticInteractionsRadical} as well as diverse and tunable magnetic ground states in triangulene-based structures \cite{Li2020,MiBeEi.20.CollectiveAllCarbon,Mishra2021,Hieulle2021}.

Another class of nanographene structures that has recently attracted interest for its unconventional magnetic properties is that of oligo(indenoindene) (OInIn), consisting of alternating hexagons and pentagons~\cite{Majzik2018, DiGiovannantonio2020, Mishra2024, Khandarkhaeva2024, Ortiz2023, ortiz2025}.
The inclusion of pentagon rings makes the lattice non-bipartite, and Lieb's theorem \cite{Lieb1989} does not apply. Depending on the pentagon-hexagon attachment orientation, different OInIn isomers can be formed, and these exhibit a radical character with unpaired electrons mainly localized at the tips of the pentagon rings \cite{Ortiz2023}. Overall, the OInIn family as a whole is a minimal yet rich platform to explore interplay between magnetic frustration, strong correlations, and non-bipartite topology. 

The quasi-one-dimensional geometry of OInIn naturally motivates the use of the density matrix renormalization group (DMRG) algorithm based on matrix product states (MPS)~\cite{White1992,Schollwoeck2011}, which is particularly well suited for one-dimensional systems whose low-energy states exhibit limited entanglement. In the context of nanographenes, DMRG has mostly been employed at the level of effective spin models~\cite{Mishra2021,Zhao2024, ZhYaHe.25.Spinexcitationsnanographene, henriques2025predictiontopologicalphasetransition, delcastillo2025remotespincontrolhaldane}.
Applications to explicit electronic Hamiltonians remain comparatively scarce: Ref.~\cite{Catarina2022} studied an effective 4-site electronic model representing each triangulene in a chain geometry,
while Ref.~\cite{CaTuKr.24.ConformationalTuningMagnetic} addressed the full $\pi$-electron description of a twisted phenalenyl dimer. Extending such electronic-structure DMRG approaches to larger nanographene systems with complex electronic structure therefore remains an open challenge.

In this work, we focus on a regular OInIn ladder, illustrated in \Figref{fig:spectra}(a), which has been predicted to host frustrated magnetic interactions \cite{Ortiz2023}. 
We apply DMRG to the full Fermi-Hubbard model to characterize the low-energy ground and excited states and to extract their emergent magnetic degrees of freedom beyond mean-field and small-active-space treatments. This allows us to firmly establish a quantitatively accurate connection between the electronic model and an emergent effective spin description. By optimizing different ans\"atze for delocalized spins, we show that a simple Heisenberg-type Hamiltonian can faithfully capture the magnetic properties of the ladder, with a flexible trade-off between accuracy of the prediction and simplicity of the ansatz. Beyond providing a compact and transparent description of the system, our approach clarifies how effective spin models emerge from correlated $\pi$-electron Hamiltonians in non-bipartite nanographene ladders.

The manuscript is organized as follows. In \Secref{sec:model} we introduce the Fermi-Hubbard description of the OInIn ladder and the corresponding $J_1$--$J_2$ spin-chain model. In \Secref{sec:spectral-matching} we compare the low-energy spectra of both descriptions and determine the spin-chain parameters that best reproduce the Hubbard-model results. In \Secref{sec:deloc-spins} we construct optimized delocalized fermionic modes that act as effective spin degrees of freedom. In \Secref{sec:quantmag} we benchmark this mapping by analyzing magnetic observables and correlations in terms of these modes. Finally, \Secref{sec:conclusions} summarizes our conclusions. Additional technical details and supplementary analyses are provided in the Appendices.

\section{Model}
\label{sec:model}

\begin{figure}
    \centering
    \includegraphics[width=\linewidth]{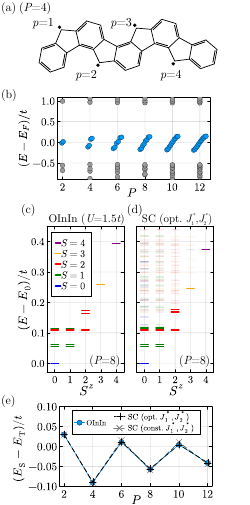}
    \caption{(a) Schematic of a regular oligo(indenoindene) (OInIn), made up of $P$ pentagons and $P+1$ hexagons, with an unpaired electron at each pentagon in the representation with the maximum Clar sextets (one at each hexagon). (b) The single-particle spectra of OInIn for different $P$, with the $P$ quasi-zero modes (offset in the $x$ axis for better visualization) highlighted in blue. (c) Parts of the energy spectra (positive $S^z$ only) of the $P=8$ OInIn chain (62 sites), and (d) the 8-spin spin chain (SC). The highlighted states are those corresponding to panel (c). (e) Difference between the lowest energy singlet and lowest energy triplet for various system sizes, for the OInIn ladder (blue circles) and the SC with parameters optimized for each $P$ (black ``$+$'' markers) and with fixed optimal $P=4$ parameters $J_1^{*}=-0.06$, $J_2^{*}=0.26$ (gray ``$\times$'' markers).
    }
    \label{fig:spectra}
\end{figure}

We describe the $\pi$-electron system of OInIn isomers by the Fermi-Hubbard Hamiltonian
\begin{equation}
\label{eq:FH_hamiltonian}
    H = -t\sum_{\langle i,j \rangle, \sigma}c^{\dagger}_{i,\sigma}c_{j,\sigma}+U\sum_{i}n_{i,\uparrow}n_{i,\downarrow},
\end{equation}
where $c_{i,\sigma}^{(\dagger)}$ destroys (creates) an electron of spin orientation $\sigma$ at site $i$, and $n_{i,\sigma}=c_{i, \sigma}^{\dagger}c_{i, \sigma}$ is the associated number operator, with $\langle i,j\rangle$ indicating the sum over nearest neighbors on the lattice.
The model parameters $t$ and $U$ correspond to hopping amplitude and on-site Coulomb repulsion, respectively.
While quantum-magnetic phases of the Hubbard model are usually associated with the large-$U$ regime, where interactions suppress the double occupation of sites, the geometric properties of OInIn give rise to quantum magnetism even for weak $U$. The key lies in its single-particle energy spectrum ($U=0$), which exhibits a relatively flat gapped manifold of $P$ spatial modes around the (mid-gap) Fermi energy $E_F$, see blue symbols in \Figref{fig:spectra}(b), with $P$ being the number of pentagon rings in the system. At half filling, the weak kinetic energy of these modes facilitates their single occupation, effectively forming a Mott insulator. This allows for a mapping onto a spin-1/2 model, in which the finite spatial overlap of the hybridized modes gives rise to a ferromagnetic (FM) exchange term, stemming from the Hubbard interaction. In addition to this, antiferromagnetic (AFM) superexchange interactions stem from second-order hopping processes. The competition of both contributions results in an effective $J_1\text{--}J_2$ Heisenberg model with one spin per pentagon, where $J_1<0$ are FM nearest-neighbor couplings, and $J_2>0$ AFM next-nearest-neighbor couplings:
\begin{equation}
\label{eq:eff_Hamiltonian}
    H_{\text{SC}} = J_{1} \sum_{p=1}^{P-1}\bm{S}_{p}\cdot\bm{S}_{p+1} + J_{2} \sum_{p=1}^{P-2}\bm{S}_{p}\cdot\bm{S}_{p+2}.
\end{equation}
The correspondence between the OInIn Hubbard model and the $J_1\text{--}J_2$ Heisenberg model has been developed in Ref.~\cite{Ortiz2023}, based on an analysis via mean-field and complete-active-space methods, showing that the ground state of the chain behaves similarly to the ground state of the target system with $J_2/|J_1|>1/4$. In this regime, a small but finite spin gap stabilizes a valence bond solid (VBS) of $S=1$ dimers, often called the Haldane dimer phase, with Affleck-Kennedy-Lieb-Tasaki (AKLT) type topological order~\cite{Bursill1995, Furukawa2012, Agrapidis2019}.

\begin{figure*}
    \centering
    \includegraphics[width=\linewidth]{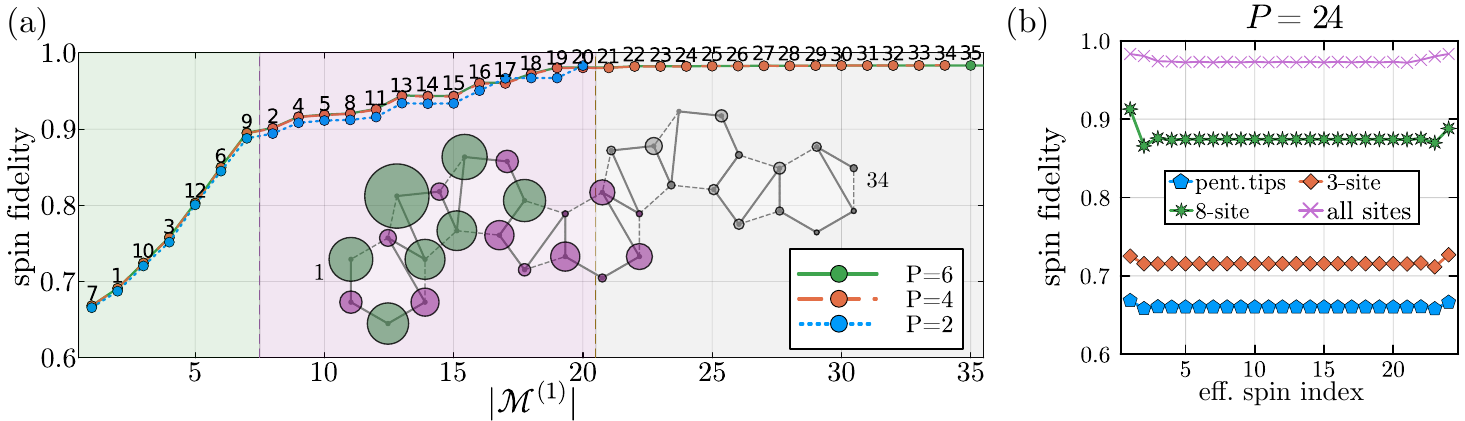}
    \caption{(a) Maximum spin fidelity of the leftmost ($p=1$) effective mode versus the number of lattice sites used to span it for $P=2$ (blue), $P=4$ (orange) and $P=6$ (green) OInIn ladders. The size of each circle on the $P=4$ chain in the inset is proportional to the magnitude of the corresponding optimized component of $\bm{\alpha}^*$. The spin fidelity systematically increases by considering more sites in the set $\mathcal{M}^{(p=1)}$, starting from the pentagon tip, where the number over each point denotes the new site index that is added, in MPS order (see inset). The sites are grouped into three different groups, depending on their relevance in describing the effective mode. (b) Spin fidelities obtained  for four different effective mode descriptions (see \Appref{app:othermodes}) by reusing and transferring the optimized effective modes for $P=4$ to an eigenstate of a much larger system of 24 pentagons.}
    \label{fig:modes}
\end{figure*}

\section{Spectral matching}
\label{sec:spectral-matching}
We first consider the respective energy spectra of both models. 
Here, the parameters of the Fermi-Hubbard Hamiltonian are set to $U/t=1.5$, which are in the range regarded as representative of nanographenes' physical properties~\cite{Yazyev2010, SchulerPRL2013}.
In \Appref{app:varyingU} results for other values of $U/t$ can be found.
We utilize DMRG in the form of the ITensor package~\cite{ITensor1} to obtain the low-energy part of the spectra of \Eqref{eq:FH_hamiltonian} for OInIn ladders of various sizes at half filling, and then tune the parameters of the spin chain in \Eqref{eq:eff_Hamiltonian} to find the best match for the energy gaps of the ladder for each size.
We show a side-by-side comparison of the energy levels of the $P=8$ OInIn with 62 sites (DMRG) in \Figref{fig:spectra}(c), and the 8-spin Heisenberg chain (exact diagonalization) in \Figref{fig:spectra}(d). Notably, the energy gaps of the much simpler spin chain are in good agreement with the gaps of the OInIn, and the spin multiplicities fully match. Particularly, we also find that the total spin of the ground state and first excited state switches between singlet and triplet depending on whether $P/2$ is even or odd, a characteristic feature of \Eqref{eq:eff_Hamiltonian} within the parameter range $J_2/|J_1|>1/4$ \footnote{This behavior can be understood from the limiting case $J_2/|J_1|\rightarrow\infty$: For $P$ a multiple of four, the chain splits into two spin chains with even numbers of spins (both with singlet ground states), whereas for even non-multiples of four, it splits into two chains with odd number of spins, each with a doublet ground state. Due to the weak FM coupling, the two doublets split into a lower triplet and a slightly higher-energy singlet.}, reflected in \Figref{fig:spectra}(e) showing sign changes in the energy difference between the lowest singlet and triplet states. We find that, although the optimal values of the spin chain parameters change slightly with $P$, reusing optimal parameters across systems still gives good agreements (see \Appref{app:optimalparams}). The entanglement entropies of the OInIn eigenstates also show remarkable signatures of the underlying effective spin system (see \Appref{app:entropy}).

The Clar sextet representation \cite{Cl.72.AromaticSextet} of the OInIn ladder suggests that the spin physics predominantly originates from unpaired electrons located at the pentagon tips. To confirm this, we quantify the spin localization by calculating the projection onto the singly occupied sector of site $i$, given by $\braket{4(S^z_i)^2}=\braket{(n_{i,\uparrow}-n_{i,\downarrow})^2}$. 
We consistently find that the pentagon tip sites exhibit projections of around $\braket{4(S^z_{7p})^2} \approx 0.66$ (in our MPS order every 7th site is a pentagon tip; thus, the $p$th pentagon tip corresponds to site index $7p$, see inset of \Figref{fig:modes}(a)), higher than the other sites which range from 0.57 to 0.61. 
Although this clearly singles out the pentagon sites, it also indicates that a mapping between the electron spins at pentagon tips and the effective spins has limited predictive power, as all spin-related observables are directly limited by this quantity.
In the following, we introduce a delocalized spin mapping that can vastly improve the figures of merit introduced in the present discussion.

\section{Delocalized spins}
\label{sec:deloc-spins}
For each effective spin $p$, we define a delocalized Fermi mode on a set of sites 
$\mathcal{M}^{(p)}$ in the ladder
\begin{equation}
\label{eq:cm_def}
c_{\mathcal{M}^{(p)},\sigma}=\sum_{i\in \mathcal{M}^{(p)}}\alpha_{i}^{(p)} c_{i,\sigma},
\end{equation}
where $\alpha_i^{(p)}\in \mathbb{C}$ and $\sum_{i\in \mathcal{M}^{(p)}}|\alpha_i^{(p)}|^2=1$, such that $\{c^{\dagger}_{\mathcal{M}^{(p)},\sigma},c_{\mathcal{M}^{(p)},\sigma'}\}=\delta_{\sigma,\sigma'}$, and $\{c^{(\dagger)}_{\mathcal{M}^{(p)},\sigma},c^{(\dagger)}_{\mathcal{M}^{(p)},\sigma}\}=0$.
Each such mode can host up to two fermions, with its corresponding spin operators being
\begin{equation}
\label{eq:effspin_operators} \bm{\mathcal{S}}_{\mathcal{M}^{(p)}} = \frac{1}{2}\sum_{\sigma, \sigma'} \bm{\tau}_{\sigma, \sigma'} c^{\dagger}_{\mathcal{M}^{(p)},\sigma} c_{\mathcal{M}^{(p)},\sigma'},
\end{equation}
where $\bm{\tau}$ is the vector of Pauli matrices.
As the primary figure of merit for evaluating the quality of such an effective spin in a certain state, we employ the single occupation of the mode, given by the expectation value of $4(\mathcal{S}_{\mathcal{M}^{(p)}}^z)^2$, which we also refer to as \textit{spin fidelity}. 
We start with sets $\mathcal{M}^{(p)}$ containing every site in the ladder, and search for normalized modes $\bm{\alpha}^{(p)}$ that maximize the spin fidelity \footnote{Although in these full optimization cases the sets $\mathcal{M}^{(p)}$ will be the same for different spins corresponding to different pentagons $p$, we still label its operators with $\mathcal{M}^{(p)}$ for short, $\mathcal{M}^{(p)}$ representing the mode with support around the $p$th pentagon tip.}.
This constrained optimization task is carried out using the NLopt package for Julia~\cite{NLopt, COBYLA}.
The procedure is remarkably well-behaved, and initializing the optimization with a mode localized at the tip of pentagon $p$, i.e., $\alpha_i^{(p)}=\delta_{i,7p}$ consistently yields the optimized mode that corresponds to the $p$th effective spin. 
These modes show robust and recurring spatial structures, independent of the system size, illustrated in the inset of \Figref{fig:modes}(a) for $P=4$. Around 40\% of the weight is on the pentagon tip, and the rest is predominantly distributed on the 6 sites highlighted in green (three in each of the neighboring hexagons), which together produce a mode with spin fidelity of 0.9.
Including also the purple sites, comprising the remaining sites in the hexagon rings around the $p$th pentagon tip and a few more remote sites, the spin fidelity reaches $0.98$. All the remaining gray sites together only increase the spin fidelity by 0.003, indicating that they play a more negligible role in the mode’s structure.

The increase of spin fidelity as a function of the size of $\mathcal{M}^{(p)}$ is systematically illustrated in \Figref{fig:modes}(a) for the leftmost mode $p=1$. 
We emphasize that for the three different system sizes shown in this plot the optimized modes almost perfectly agree, highlighting a potent feature: Independent of the system size, the optimization can be performed on small sets $\mathcal{M}^{(p)}$ that only contain the relevant (green and purple) sites. In this way, the number of variables to optimize over remains constant, and the complexity of evaluating the cost function scales favorably with $P$ \footnote{Only optimizing over the green and purple sites means that the single occupation operator only spans a few tensors. With an appropriate gauge transformation of the MPS, the number of bonds to contract over remains constant with $P$.}. Moreover, once the optimized modes have been obtained for one system size, one may even fully avoid further optimization procedures, as modes can be reliably transferred between systems of different sizes. With suitable index permutations and parity considerations, the optimal modes are very similar across different effective spins within a given chain, allowing the characterization of all effective modes in a larger chain using the modes from a smaller one. This can be seen in the purple crosses of
\Figref{fig:modes}(b), where the four optimal modes for $P=4$ were used to construct all 24 modes of a $P=24$ ladder, still achieving spin fidelities close to 1. 
Finally, we also emphasize that the optimized modes are largely insensitive to the chosen molecular eigenstate. However, we note that the achieved spin fidelity decreases when eigenstates with smaller total spin $S$ are considered. For instance, at $P=6$ we obtain spin fidelity of 0.98 in the $S=3$ sector, but this value decreases to 0.92 for states with $S=2$, 0.84 for $S=1$, and 0.81 for $S=0$. This is expected for $P$ electrons in $P$ effective modes (see \Appref{app:singlets}).

\section{Quantum magnetism of delocalized modes}
\label{sec:quantmag}

\begin{figure*}
    \centering
    \includegraphics[width=\linewidth]{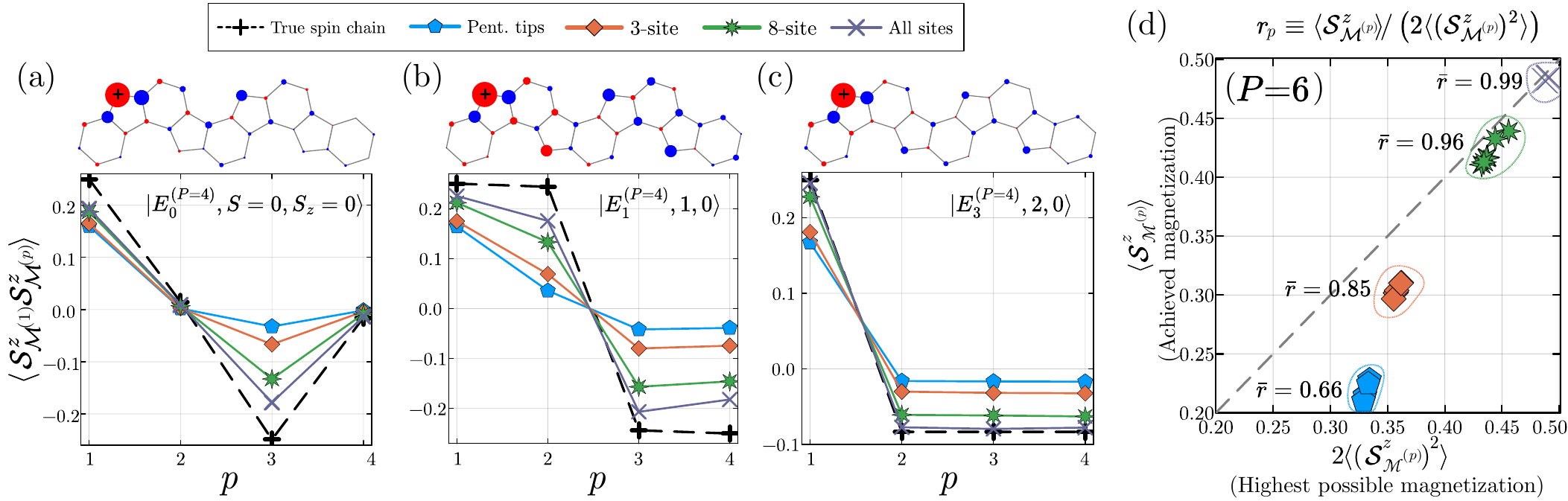}
    \caption{(a)--(c) Spin correlations in the $P=4$ system for different effective mode descriptions (pentagon tip, 3-site, 8-site, and all-site modes), compared with the target spin chain correlations (dashed black crosses). Results are shown for three eigenstates with $S^z = 0$: 
    (a) the singlet ground state, dominated by singlets between second neighbors; 
    (b) the first excited triplet, mainly composed of configurations where one of the singlets of the ground state is broken $\ket{\uparrow\uparrow\downarrow\downarrow}-\ket{\downarrow\downarrow\uparrow\uparrow}$; 
    (c) the $S^z=0$ state of the $S=2$ multiplet, corresponding to a fully symmetric Dicke state with two spins up and two spins down.
    The structures on top of each plot show the site-wise correlation values of the corresponding eigenstate, where the reference spin has been placed on the first pentagon tip.
    The area of the circles on each site is proportional to the magnitude of the correlation and the color indicates its sign.
    (d) The measured and maximum possible magnetizations for different effective mode types for all effective spins within the state lowest energy $\ket{3,3}$ state for $P=6$.
    A perfect fidelity spin that exactly mimics the target spin would lie on the top right corner of the plot. We also show the averaged magnetization approximation ratios $\bar{r}=\sum_pr_p/P$ for each mode type.
    }
    \label{fig:corrs}
\end{figure*}

Finally we benchmark the delocalized modes by verifying how accurately they mimic their counterparts in the frustrated Heisenberg chain.
We concentrate on four different types of effective modes: the localized pentagon tip mode, the fully optimized modes, as well as intermediate 3-site and 8-site modes (see \Appref{app:othermodes} for details).
In \Figref{fig:corrs}(a)--(c), we show the spin correlations of the OInIn ladder for three distinct eigenstates after mapping with the four types of effective spins, and compare with the correlations in the corresponding eigenstates of the Heisenberg chain. We find that their shapes match (exceptions are covered in \Appref{app:asymmetry}) and that the agreement becomes more accurate as we use more complex modes.

We further quantify the quality of the delocalized mode description by looking at their magnetizations. The maximum magnetization of a mode is given by half its spin fidelity, $2\braket{(\mathcal{S}_{\mathcal{M}^{(p)}}^z)^2}$. For pentagon tips, $\mathcal{M}^{(p)}=\{7p\}$, this comes to be $\approx 0.33$, significantly below 1/2. However, the magnetization of the pentagon tips remains considerably below this maximum value for all states. For instance, the lowest energy OInIn eigenstate with total spin $S=P/2$ and $z$-component $S^z=P/2$ should correspond to a fully polarized spin state $\ket{\uparrow}^{\otimes P}$, but we find the magnetization of the pentagon tips to be slightly below $0.22$.
In \Figref{fig:corrs}(d) the achieved magnetization is plotted against the maximally possible magnetization for the four different mode types for the $P=6$ state $\ket{3,3}$. Data points close to the diagonal line indicate that the singly occupied components of a mode faithfully reproduce the target state, expressed through a magnetization approximation ratio $r_p=1$.
Remarkably, even the relatively simple 3-site mode can significantly improve $r_p$, as compared to the fully localized spin map. The 8-site modes already provide a very accurate description of the spins, close to the fully optimized (all sites) modes, while being significantly less complex, with only 3 parameters to optimize over.
In addition, utilizing the fully optimized modes, with spin fidelity over 0.98 and $\bar{r}=0.99$, implies that the magnetization of the state is dominated by contributions from those $P$ modes.

Finally we look at the fidelity with which two molecular eigenstates are connected through the flip of one effective spin, shown in Tab.~\ref{tab:spinflip} for the spin flip of a fully polarized state at $P=4$. This fidelity is a crucial quantity for dynamically probing the spin picture, e.g., by applying localized fields to the OInIn via scanning tunneling microscopy (STM). Importantly, it can be increased from 0.19 for the localized pentagon tip spins to 0.94 in the fully optimized modes.

\begin{table}
\centering
\begin{tabular}{l c c c} 
 \hline\hline
 \multirow{2}{*}{$\mathcal{M}$ type} & 
 Spin & 
 \multirow{2}{*}{Magnetization} &
 Spin-flip \\
  & fidelity &  & fidelity \\ 
 \hline
 \pentagofill[cyan]~pent.\ tips & 0.66 & 0.22 & 0.19 \\ 
 \rhombusfill[orange]~3-site & 0.72 & 0.31 & 0.37 \\
 \eightpointedstar{Green}~8-site & 0.90 & 0.42 & 0.72 \\
 \textcolor{violet}{$\times$}~all sites & 0.98 & 0.48 & 0.94 \\
 \hline\hline
\end{tabular}

\caption{Spin fidelity, magnetization, and spin-flip fidelity, defined as
$|\braket{\frac{P}{2},\frac{P}{2}-1|\mathcal{S}^-_{\mathcal{M}}|\frac{P}{2},\frac{P}{2}}|^2$,
for the lowest-energy eigenstate with $S=S^z=P/2$ with different effective mode types.
The data refer to $P=4$, averaged over the $P$ modes, but the same values are obtained
for $S=P/2$ eigenstates for other system sizes.}
\label{tab:spinflip}
\end{table}

\section{Conclusions}
\label{sec:conclusions}
We proposed a DMRG-based framework to characterize emergent effective spin physics in quasi-1D nanographene systems, without the need to define an active space and without restrictions on the interaction parameters of the electronic Hamiltonian. Within this approach, the relevant magnetic degrees of freedom emerge naturally from the many-body optimization.
Applying it to a class of OInIn ladders, we established an accurate correspondence between the correlated electronic description and a simple $J_1$--$J_2$ spin chain in the $J_2/|J_1|>1/4$ regime capturing the low-energy magnetic properties.
However, for finite systems, AKLT-like physics can become obscured for large values of $J_2/|J_1|$ due to the small stabilizing spin gap and large correlation length \cite{Agrapidis2019}, suggesting the study of long chains which suitably tuned effective spin-chain parameters (e.g., via dielectric screening or structural modifications).
Beyond providing a compact and transparent description of frustrated nanographene ladders, our results offer a controlled starting point for further low-energy modeling, including extensions to effective $t\text{--}J$ descriptions that incorporate charge fluctuations and doping. The approach can also be applied to extended models incorporating longer-range interactions or multi-orbital effects.
Our work further provides the basis for dynamically probing the effective spins, for example via STM techniques, and for exploring carbon-based spin-chain physics, where future work may reveal novel spin behaviors.

\section*{Acknowledgments}
We thank R.\ Ortiz and G.\ Catarina for useful comments and discussion. The research was funded by the Department of Education of the Basque Government through PIBA\_2023\_1\_0021 (TENINT), by Agencia Estatal de Investigación MCIN/AEI/10.13039/501100011033 through Proyectos de Generación de Conocimiento PID2022-142308NA-I00 (EXQUSMI) and PID2023-146694NB-I00 (GRAFIQ), as well as by the European Union NextGenerationEU/PRTR-C17.I1 and by the IKUR Strategy under the collaboration agreement between Ikerbasque Foundation and DIPC on behalf of the Department of Education of the Basque Government with in project QT13: DREAMS.

\section{Data Availability}
The data that support the findings of this article are openly available~\cite{dataset2026}.

\appendix

\section{Designing other effective modes}
\label{app:othermodes}
\begin{figure}[b]
    \centering
    \includegraphics[width=\columnwidth]{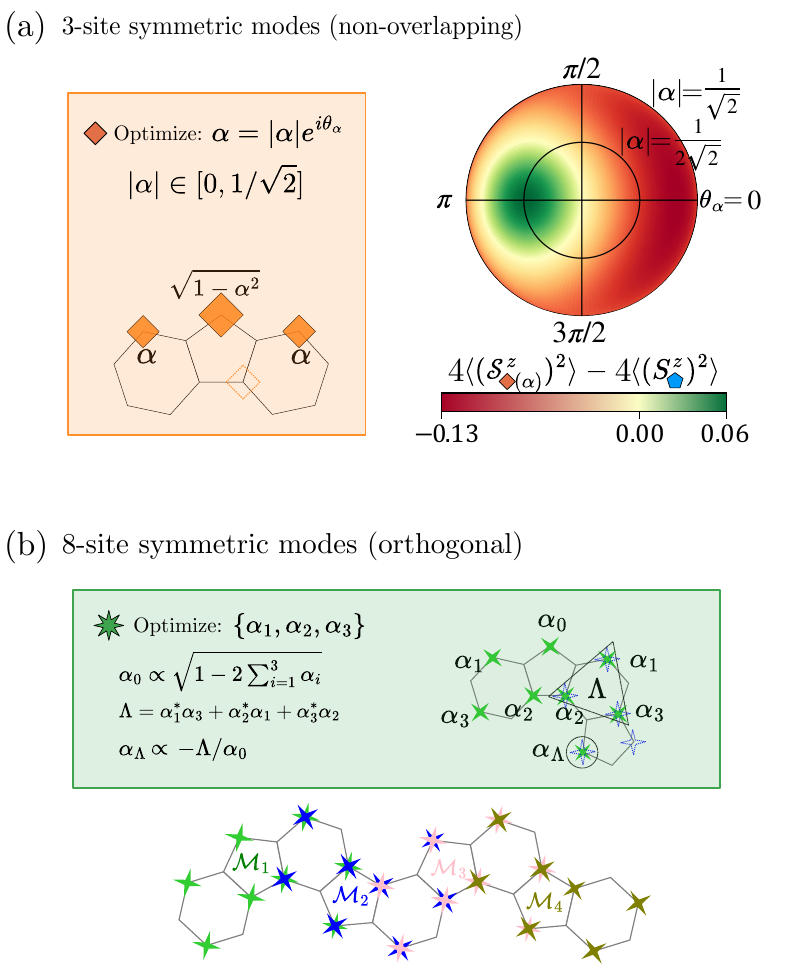}
    \caption{Schematics of 3-site and 8-site modes used throughout the main text. (a) Left: 3-site symmetric modes with one parameter ($\alpha$) to optimize over. Right: change in spin-fidelity with respect to the pentagon tip's for all values of $\alpha$. (b) 8-site symmetric modes with 3 parameters to optimize. The small weights on the next pentagon tips over exactly cancel out all overlaps.}
    \label{fig:other_modes}
\end{figure}

The observations of \Figref{fig:modes}(a) motivate the use of the regular patterns of the optimal modes for designing simpler (fewer variables to optimize, lesser spread through the ladder) effective modes. In this section we study the validity of this approach by proposing two such intermediate modes which are used throughout the main text.

In \Figref{fig:other_modes}(a) we consider a simple effective mode which in addition to the pentagon tip $p$ has support on the two sites on either site of $p$, which should give a small improvement with respect to the pentagon-tip-only description. The amplitude on either side of the pentagon is chosen to be the same, and since the amplitude at the pentagon tip is given by normalization and overall phase-fixing, this mode is controlled by one complex parameter $\alpha$. We see that when chosen optimally it provides a modest increase of 0.06 in spin fidelity with respect to the pentagon tip.

We propose another mode in \Figref{fig:other_modes}(b), which makes use of all of the important sites, that is, the ones labeled in green in \Figref{fig:modes}(a). By making the modes symmetric around the pentagon tip again, we reduce the number of complex variables to optimize over to 3, and the weight on the pentagon tip is again given by normalization and can be chosen to be real.
In this case, unlike for the 3-site mode, the modes corresponding to two neighboring effective spins overlap. If this overlap, which for two effective modes on sites $\mathcal{M}^{(p)}$ and $\mathcal{M}^{(p')}$ is defined as $\Lambda=\sum_{i\in\mathcal{M}^{(p)}\cap\mathcal{M}^{(p')}}(\alpha_i^{(p)})^*\alpha_i^{(p')}$ is non-zero, the operators for the two modes do not anticommute, and the respective spin operators do not commute and do not have a joint eigenbasis in which to describe the states. 
Because of this, we modify the mode corresponding to the spin $p<P$ by adding to the set $\mathcal{M}^{(p)}$ the next pentagon-tip site $p+1$, with an amplitude of $\alpha_\Lambda=-\Lambda/\alpha_0$, where $\alpha_0$ is the amplitude on pentagon tip $p$, such that the overlaps of these two modes are exactly canceled out (the mode corresponding to the spin $p=P$ is not modified unless one considers periodic boundary conditions). See \Figref{fig:other_modes}(b).

We note that both of these modes are universal: they are optimized only once, for one effective mode of one eigenstate of an OInIn of $P$ pentagons, and can then be transferred to the other effective spins of the ladder (by permuting the sites/amplitudes accordingly), and across different eigenstates and system sizes, as seen in \Figref{fig:modes}(b), making these effective mode descriptions highly efficient.

Many other effective spins can be experimented with, such as ones which have alternating odd/even versions, which can also model the asymmetry in the system (see \Appref{app:asymmetry}), or ones where the complex nature of the components of $\bm{\alpha}^{(p)}$ is exploited to enforce orthogonality.

\section{Effects of coupling asymmetries}
\label{app:asymmetry}
\begin{figure}
    \centering
    \includegraphics[width=\columnwidth]{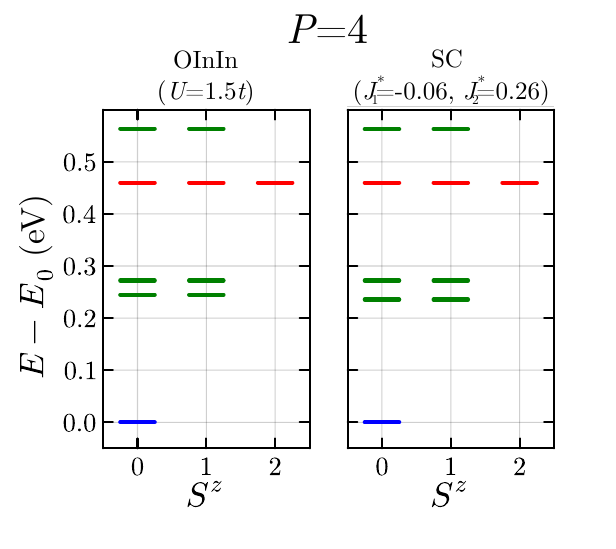}
    \caption{Spectra comparison for $P=4$. Note the slightly smaller gap between the first two triplet states for OInIn.}
    \label{fig:P4_spectrum}
\end{figure}

\begin{figure*}
    \centering
    \includegraphics[width=0.9\linewidth]{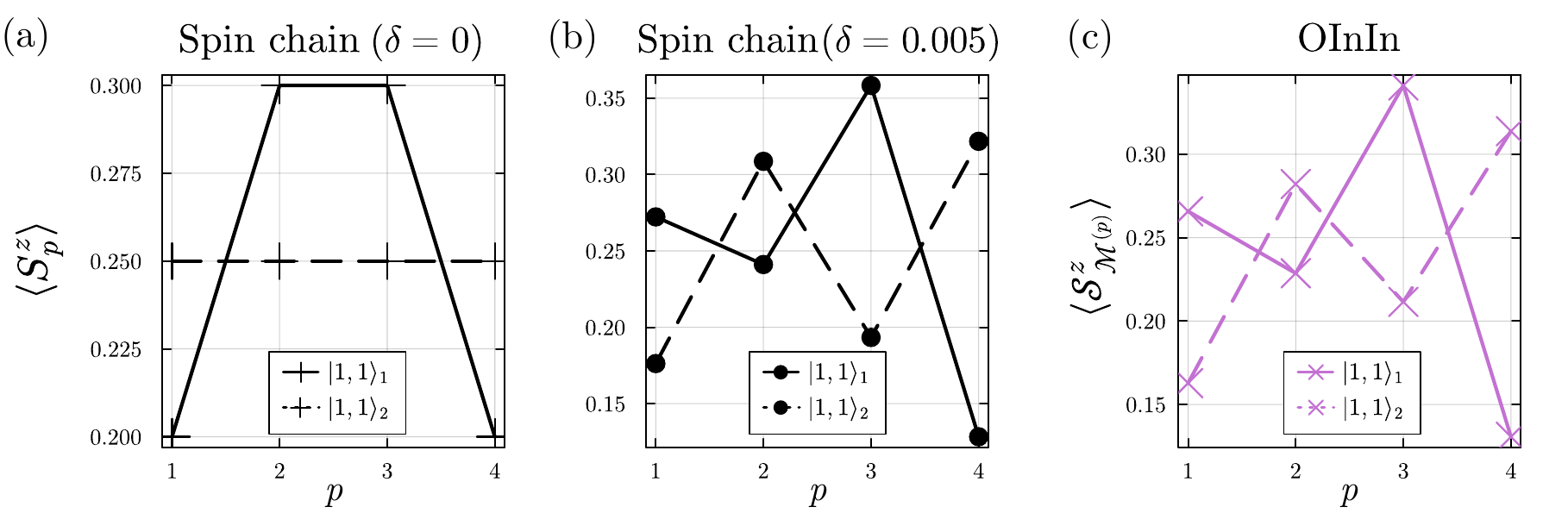}
    \caption{The magnetization profiles of the $P=4$ ground (solid lines) and first excited state (dashed lines) for the $S^z=1$ sector of (a) the uniform spin chain, (b) symmetry-broken spin chain, (c) the OInIn ladder with optimized (All sites) modes. Adding an asymmetry term mixes the two triplet states and qualitatively matches the magnetizations. This suggests that the spin chain couplings are not uniform across the chain. Notice, that $\delta$ is two orders of magnitude smaller than $J_2^*$, meaning that only a few specific states can be mixed this way.}
    \label{fig:asymmetry}
\end{figure*}
The source of the effective spin interactions between the delocalized modes are the different exchange processes mediated by the on-site Coulomb interaction and different hopping processes \cite{Ortiz2019, Jacob2022}. In the case of the OInIn systems, with their lack of mirror symmetry and without zero modes that can be more regularly localized like with cleaner bipartite lattices, the slight irregularities in the modes can lead to differing overlaps between the different effective spins. These give rise to slight differences in their respective couplings.

This results in no noticeable effect for the most part, and the observables match with the target system when making use of the optimized modes. However, these asymmetries can mix certain states with the same $S$ and $S^z$ quantum numbers, such as is the case with the two lowest energy triplets of OInIn with $P/2$ even. The small energy gap (see \Figref{fig:P4_spectrum}) separating these is comparable to the asymmetry generated in the couplings by the effective modes. This mixing also makes the triplet gap smaller in the OInIn spectrum compared to the spin chain.

This is very evident when attempting to match the physical observables of these eigenstates with the target spin chain. We show this discrepancy for $P=4$ in \Figref{fig:asymmetry}(a),\! (c). We may model this by adding an asymmetry term $\delta_i$ for $i=1,2$ to the spin chain parameters $J_1, J_2$, respectively. We show in \Figref{fig:asymmetry}(b) that by introducing a nonzero $\delta_2$,  $H'=\delta(\bm{S}_1\cdot\bm{S}_3-\bm{S}_2\cdot\bm{S}_4)$, which modifies the $J_2$ couplings of the chain we are able to reproduce the the OInIn model. Note that the asymmetry value in $J_2$ is two orders of magnitude smaller than $J_2$ itself. Meaning that this only affects the spin chain picture in very specific cases.

It is important to note that this effect is clearly manifest in the physical observables for $S^z\neq0$ states only. Notice that the middle plot of \Figref{fig:corrs}(a) is one of these states with $S^z=0$, giving good agreement and showing no asymmetry effects.

\section{Varying the on-site interaction}
\label{app:varyingU}
\begin{figure*}
    \centering
    \includegraphics[width=\linewidth]{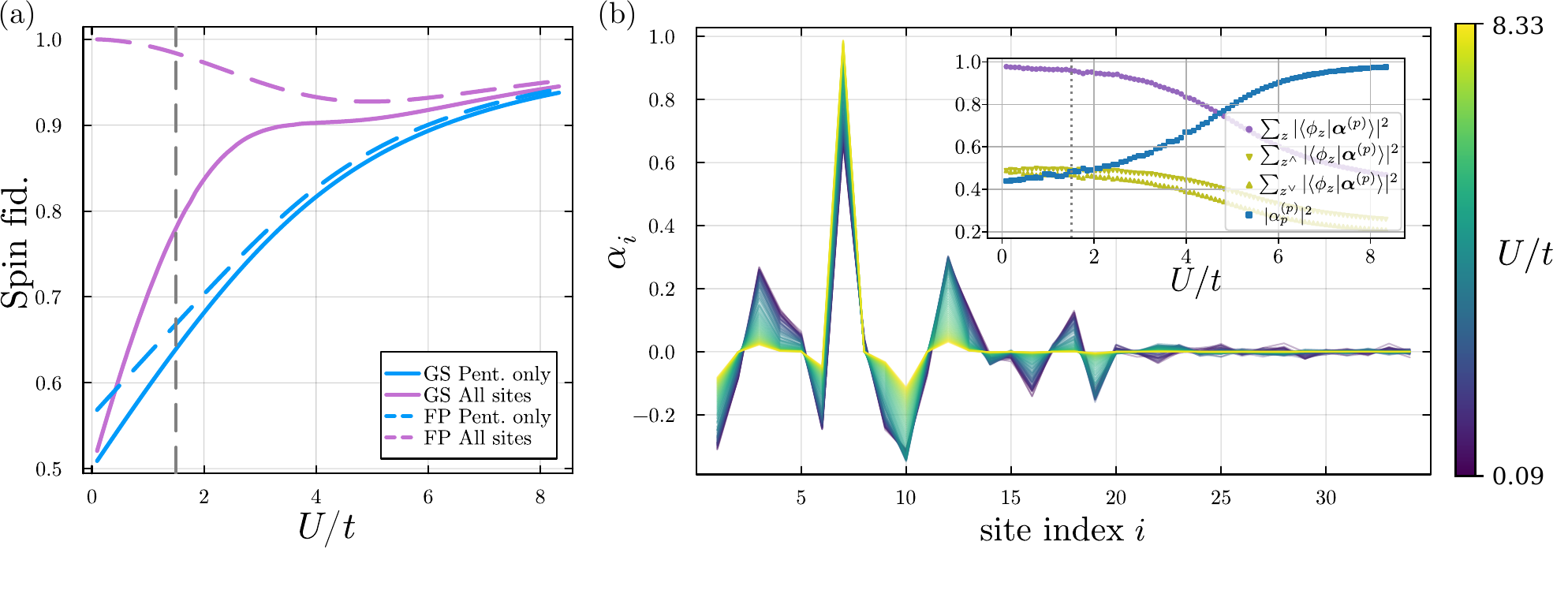}
    \caption{(a) The spin fidelities of the pentagon-tip-only mode (blue) and the fully optimized mode (purple) as a function of on-site interaction for the ground state (GS) of OInIn (solid) and the state the equivalent of which is the fully polarized (FP) state of the spin chain (dashed). The mode is optimized for every $U$ value. The gray vertical line indicates $U/t=1.5$. (b) The optimized mode for the $P=4$ ground state for different $U$ values. The site indices on the $x$ axis are in MPS order (see \Figref{fig:modes}(a)). Inset: the overlap of the optimized mode $\bm{\alpha}^{(p)}(U)$ with the space of the $P$ quasi-zero modes $\ket{\phi_z}$, $z\in\{1,\dots,P\}$, (purple circles), the $P/2$ quasi-zero modes, $z\in\{1,\dots,P/2\}$, with energy $E<E_F$ (green up triangles) and $E>E_F$, $z\in\{P/2,\dots,P\}$, (green down triangles). The blue squares represent the weight on the pentagon-tip site of the fully optimized mode.}
    \label{fig:spm_analysis}
\end{figure*}

In this section we explore the effect of varying the on-site interaction parameter of the Fermi-Hubbard Hamiltonian on the effective-mode picture. To this end, we consider a $P=4$ ladder with  different $U$ values and calculate for each value the optimized effective mode that maximizes the spin fidelity. We do this both for the $\ket{2,2}$ state, equivalent to the fully polarized (FP) state in the effective spin model, and the $\ket{0,0}$ singlet ground state (GS), both obtained with DMRG.

In \Figref{fig:spm_analysis}(a) we plot the spin fidelities of the fully optimized modes (purple) and the pentagon tips (blue) as a function of $U/t$ for the GS (solid lines) and FP (dashed lines) state. The pentagon-tip spin fidelities show similar behavior for both eigenstates, their values increasing with $U/t$ as the double occupancies are increasingly penalized, reaching exact single occupation in the limit of strong interaction to form a Heisenberg model on the OInIn lattice, where $N$ spins are fully localized on the $N$ individual lattice sites. Indeed, all lines eventually converge in the strongly interacting limit, as the optimized mode gets closer and closer to occupying just the pentagon-tip site itself. This can be seen in \Figref{fig:spm_analysis}(b), where the amplitudes of all of the optimized modes for different $U/t$ are plotted. At high $U/t$, the amplitude at the pentagon-tip site ($i=7$) approaches one and all other sites' amplitudes either vanish or significantly decrease. 

Turning to the fully optimized modes, we observe in \Figref{fig:spm_analysis}(a) that for small $U/t$ these behave very differently in the two eigenstates. In the FP case, we obtain perfect spin fidelity, since enforcing $S^z=P/2$ restricts the $P$ quasi-zero modes (see \Figref{fig:spectra}(b)) to the only possible configuration that yields the required magnetization, where all of these modes are perfectly singly occupied. Therefore, the spin fidelity starts to drop as we increase the interaction strength as other single-particle modes start fluctuating in particle number. Indeed, the deviation of the spin fidelity from 1 for the fully polarized state of $P$ effective spins is a direct indication of a limitation of CAS$(P,P)$, which necessarily yields perfect spin fidelity. In the case of the GS, the effects of geometric frustration (which causes the quasi-zero modes to deviate from exact zero energy) are manifest and a completely different behavior is seen for the optimized mode in the weakly interacting limit: the fact that the $P$ modes that function as the active orbitals at low interaction are not real zero modes means that $U$ needs to overcome that initial energy-barrier in order to be able to hybridize the modes, and thus, the spin fidelity of the mode starts low and increases with $U$. In the inset of \Figref{fig:spm_analysis}(b) we observe that the optimized modes for small $U/t$ can be fairly accurately expressed as linear combinations of these $P$ modes (purple circles).
The $U/t\approx2$ value where the difference between the spin fidelity of the pentagon-tip site and that of the fully optimized mode is largest can be used as an estimate for the optimal $U$ value, where the $P$-effective-spins picture is the strongest.

In the limit $U/t\rightarrow 0$ the optimization procedure yields modes with spin fidelity equal to 1/2. This can be understood as follows: in a closed-shell ground state, the first $N/2$ energy eigenmodes are doubly occupied, and the rest are empty.
Any normalized singe-particle mode $\alpha$ can then be written as a linear combination $\alpha=\cos(\theta) f + \sin(\theta) e$, where $f$ and $e$ are modes in the doubly filled and empty sector, respectively. Then a quick calculation shows that $\langle 4(S^z)^2\rangle_\alpha = 2\cos(\theta)^2(1-\cos(\theta)^2)$ reaches its maximum for $\cos(\theta)^2=1/2$, yielding an optimal spin fidelity of 1/2 for the non-interacting system. This can be confirmed from the inset of \Figref{fig:spm_analysis}(b), where the overlaps with the subspaces of the single-particle modes that at $U=0$ are doubly occupied and empty are plotted separately ($\ket{\phi_z}$, where $z \in \{1,\dots,P/2\}$ and $z \in \{P/2,\dots,P\}$, respectively), confirming that they become equal for small $U/t$ in order to obtain a mode that maximizes single occupation, with a single occupation of 1/2.

\section{Matching the entanglement entropy}
\label{app:entropy}

\begin{figure*}
    \centering
    \includegraphics[width=0.9\linewidth]{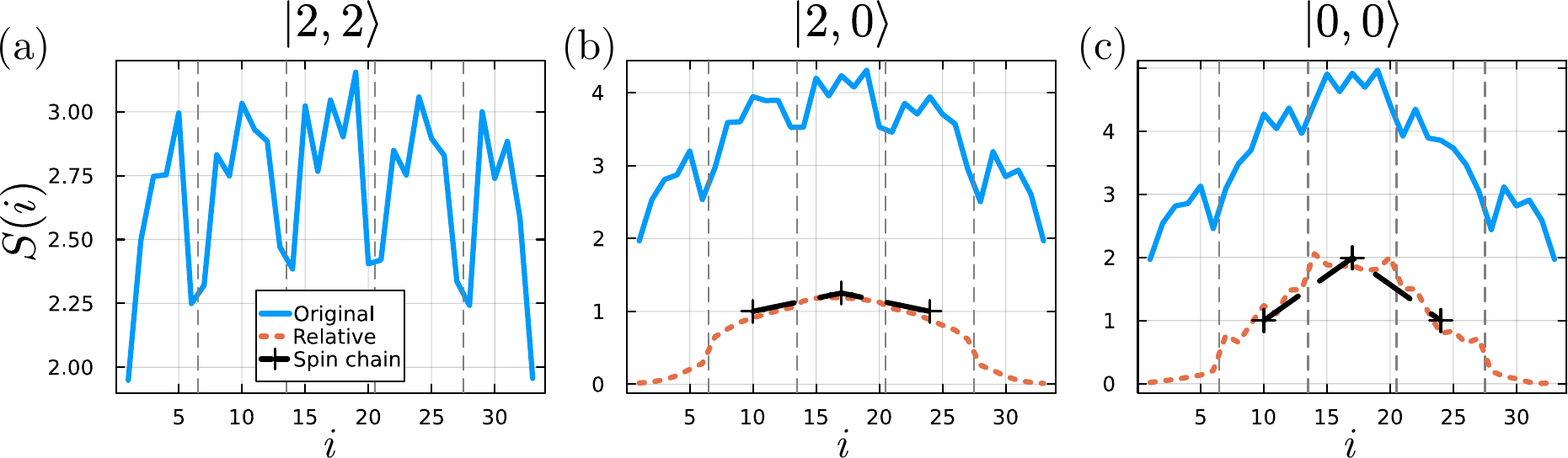}
    \caption{Entanglement entropy across bipartitions at different bonds $i$ in the MPS for $P=4$, $S(i)\equiv S(1, ..., i|i+1, ..., N)$, (a) for the reference eigenstate, the equivalent of which in the spin chain is the unentangled fully polarized state $\ket{\uparrow\uparrow\uparrow\uparrow}$, (b) for an excited state and (c) ground state of the same OInIn, including the original entanglement profile (blue) the difference from the reference entropy of the plot in (a) (orange), as well as the entropy of the equivalent spin chain eigenstates (black). The four vertical dashed lines in each plot indicate the sites corresponding to the pentagon tips. The black ``$+$'' markers for the spin chain entropy were placed at the midpoints in between pentagon tips.}
    \label{fig:entropies}
\end{figure*}
The entanglement entropy with respect to a bipartition of a state (which can be straightforwardly obtained for contiguous bipartitions of MPS tensors) can also give some hints as to the existence of effective spins. In \Figref{fig:entropies}(a) we observe for the $\ket{2,2}$ state for $P=4$ a fairly constant entanglement profile with four (one per pentagon) large and distinct dips in the vicinities of the pentagon tips. Other eigenstates also show similar dips, although they are much less pronounced, with a noticeable overall increase of entropy towards the middle of the chain, than the states which correspond to product-state equivalents in the effective model. This motivates using the entropy profile of \Figref{fig:entropies}(a) as a sort of ``background'' entropy, on top of which the entropies of the target system are built. We illustrate this for the ground state of the $P=4$ ladder in \Figref{fig:entropies}(c), as well as for an excited state in (b), where we show the original entropy profile (blue) and its difference from the ``background'' entropy (orange), which shows a qualitative match to the entropy of the ground state of the target spin chain (black) \footnote{Note that the placement of the spin chain markers is not precise due to the overlap of different effective spins.}.

\section{Singlet formation and spin compatibility}
\label{app:singlets}

\begin{figure*}
    \centering
    \includegraphics[width=0.9\linewidth]{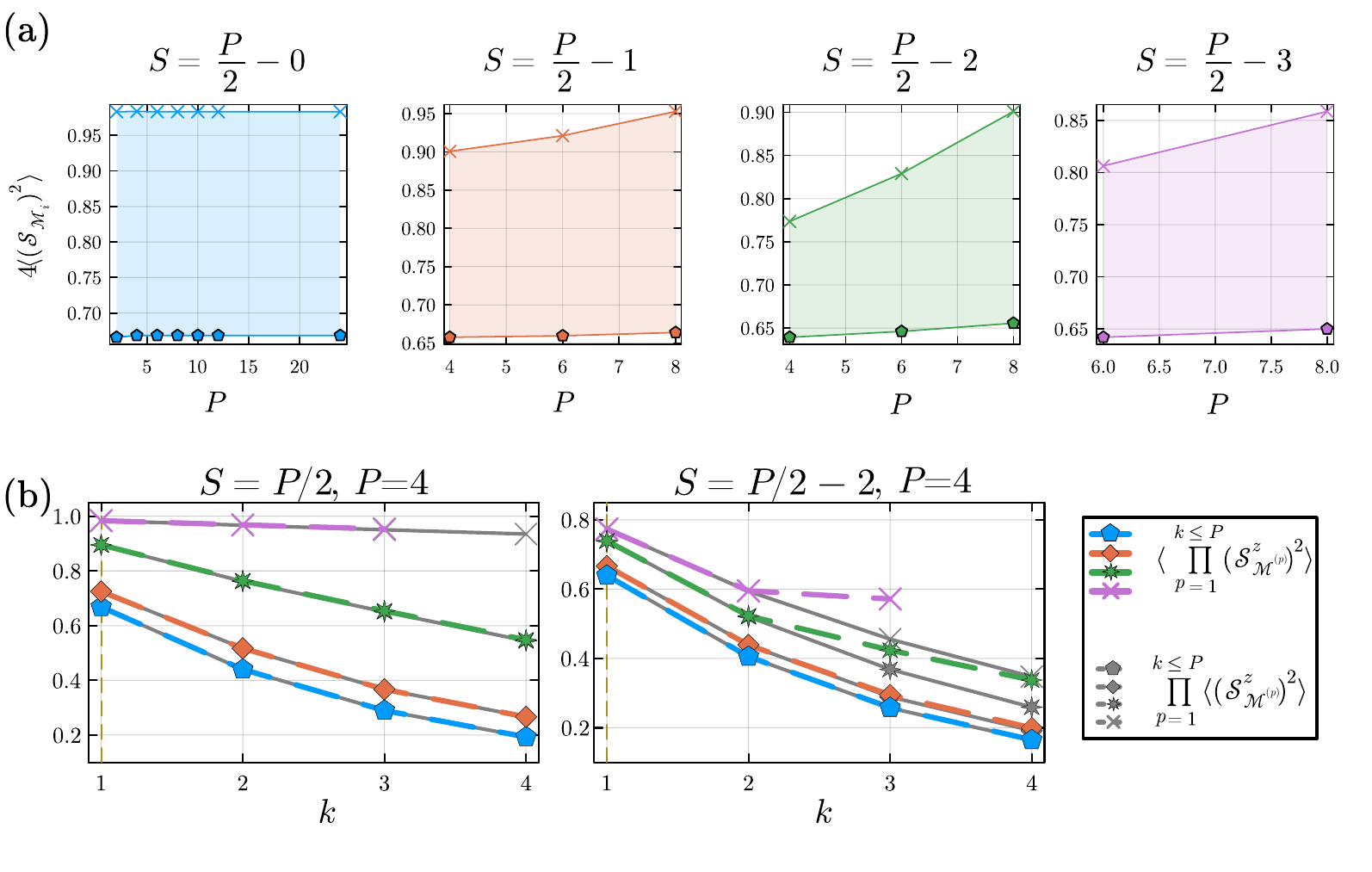}
    \caption{(a) The minimum (pentagon tips only) and maximum (fully optimized over all sites, transferred from $P=4$) spin fidelities as a function of different system sizes. (b) Spin fidelities after $k\leq P$ consecutive projections onto the singly occupied parts of the first $k$ spins of a $P=4$ chain (solid lines) for the different effective mode types, and the products of the individual spin fidelities of the first $k$ effective modes (dashed lines) for the different effective mode types. The panel on the left refers to the $\ket{2,2}$ state and the one on the right to the ground state $\ket{0,0}$.}
    \label{fig:spin_compatibility}
\end{figure*}

Singlet formation affects the spin fidelities of the effective modes. While the states of OInIn the equivalents of which are in the maximal multiplet $S=P/2$ of Dicke states exhibit effective modes with spin fidelities close to 1, those same modes for other eigenstates see their spin fidelities progressively reduced with the total spin quantum number. This effect of singlet formation between the delocalized modes is illustrated in \Figref{fig:spin_compatibility}(a), where the spin fidelity of the leftmost mode ($p=1$) and that of the pentagon tip are plotted against system size for different total spin quantum numbers. This gives a range between the best and worst possible spins for a given eigenstate.
For the maximally polarized multiplets $S=P/2$, we see a constant value of 0.98 regardless of system size as expected. As we go to lower spin multiplets, the maximum spin fidelity consistently drops, with the smaller system sizes showing the largest drop for the same integer lowering of the total spin from the $P/2$ multiplet.

Even when two high-fidelity effective modes are devised, it could still be the case that using them jointly does not provide a high-fidelity description. Specifically, if the projection of the many-body eigenstate onto the subspace where one mode is singly occupied, in turn, lowers the spin fidelity of another mode. This is relevant for considering expectation values of higher-order spin operators. To quantify this we may define the amplitude after $k$ consecutive projections onto the singly occupied sectors of different effective spins: $\braket{\prod_{p=1}^{k\leq P}(\mathcal{S}^{z}_{\mathcal{M}^{(p)}})^2}$, which can be understood as an order-$k$ spin fidelity, or a measure of simultaneous single occupation of $k$ different effective modes. We compare this to the product of their individual spin fidelities $\prod_{p=1}^{k\leq P}\braket{(\mathcal{S}^{z}_{\mathcal{M}^{(p)}})^2}$ in \Figref{fig:spin_compatibility}(b).

These two values being equal would mean that the spin fidelity of different spins is not affected after the projection. This is, in fact, the case for all eigenstates in the $S=P/2$ multiplets, which show similar values for both quantities. Since the modes in these states have very high fidelity, there is a clear advantage of using the improved mode picture, as the spin fidelity still remains high after a few projections. Looking at these quantities for other eigenstates, however, such as the $P=4$ singlet shown in \Figref{fig:spin_compatibility}(b), shows a remarkable behavior. Although the simpler pentagon-tip and 3-site modes have similar behavior as to the fully polarized cases, where they overlap with the products of the expectation values of the individual projectors, the more accurate modes (8-site and All sites) show significantly higher simultaneous fidelity (around 30\% higher) for high order $k>2$ projections compared to the (lower) values obtained from the individual spin fidelities of the modes. This means that these upgraded modes make the effective spins more compatible with each other, and a projection onto an effective mode's singly-occupied subspace implies, at the same time, an enhanced single occupation for the other modes. This could prove to be beneficial in calculations higher-order observables of interest in the target system, such as string order parameters and four-point functions.

\section{DMRG and spectral matching details}
\label{app:optimalparams}
\begin{figure*}
    \centering
    \includegraphics[width=0.9\linewidth]{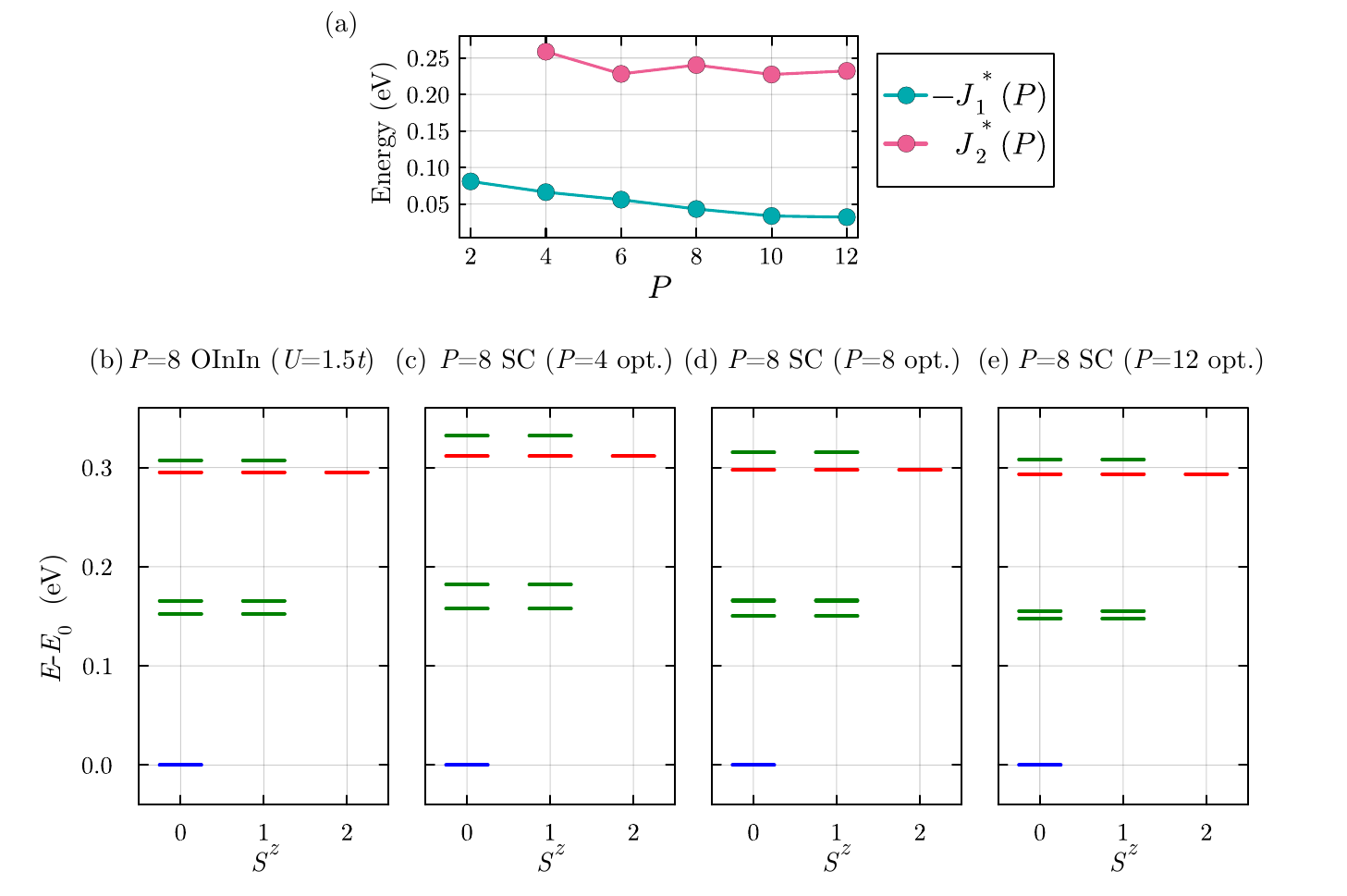}
    \caption{(a) Optimal spin chain (SC) parameters that best match the low-lying energy gaps for various $P$. Panel (b) shows the energy levels of $P=8$ OInIn, and (c)-(e) show the same energy levels for the SC with 8 spins, but with the optimal $J_1^*(P), J_2^*(P)$ values for $P=4$, $P=8$ and $P=12$. The square distances from the gaps of the OInIn ladder for the five low-lying eigenstates are $1.1\times10^{-3}$, $7.4\times10^{-5}$ and $8.7\times 10^{-5}$, respectively.}
    \label{fig:param_optimization}
\end{figure*}

The DMRG calculations were performed with a truncation threshold of $10^{-9}$ for discarding Schmidt values and with no imposed limit on the bond dimension, which was systematically increased until an energy convergence of $10^{-6}$ was achieved. In many cases increasing the number Lanczos iterations was also necessary for escaping local minima, especially for converging to the first two excited triplet eigenstates for systems with singlet ground states, which are often close in energy (see \Appref{app:asymmetry}). We know these states are well distinguished when the matrix element of the Hamiltonian in Eq.~\eqref{eq:FH_hamiltonian} connecting the two triplet states is much smaller than the energy gap between them.
Convergence in energies and entanglement entropies is typically more costly and comes after convergence in local observables is achieved. Bond dimensions of the order of $10^4$ were needed for the larger systems that we considered, and this requirement increased with each subsequent excited state. 

Once a number of energy eigenstates of OInIn were obtained via DMRG, we optimize the effective spin chain (SC) parameters by finding the ones that give the energy spectrum that minimizes the square distance between the respective gaps of the system. 
\Figref{fig:param_optimization}(a) shows the optimal SC parameters $J_1^*$, $J_2^*$ for each system size, obtained in this way.
We find that while $J_2^*$ stays mostly constant and around 0.23, $J_1^*$ seems to decrease slowly with $P$. In \Figref{fig:param_optimization}(b)-(e) we show that reusing other optimal parameters for other $P$ results in qualitatively similar energy levels (similar gaps and matching total spin ordering). As expected, the SC parameters optimized for $P=8$ give the best match, but the $P=12$ parameters only result in a very slightly worse match, and the $P=4$ ones give a noticeably worse one. This suggests that as we increase $P$ the parameters get more and more stable. It is worth noting that the regimes of the optimal SC parameters have large ratios $J_2/|J_1|$, and these values being far from the ratio of 0.6 giving the spin gap maximum means that the topological signatures of the GS are expected to be delocalized due to a much larger correlation length in this regime, their observation requiring much larger system sizes~\cite{Agrapidis2019}.

\providecommand{\noopsort}[1]{}\providecommand{\singleletter}[1]{#1}%

\end{document}